# Thermal inertia of near-Earth asteroids and implications for the magnitude of the Yarkovsky effect

Marco Delbo[1,2], Aldo dell'Oro[1], Alan W. Harris[3], Stefano Mottola[3], Michael Mueller[3]

[1]*INAF-Oss. Astron. di Torino, via Osservatorio 20, 10025 Pino Torinese (TO), Italy*

[2]*Observatoire de la Côte d'Azur B.P. 4229, 06034 Nice Cedex 4, France*

[3]*DLR Institute of Planetary Research, Rutherfordstrasse 2, 12489 Berlin, Germany*

Running head: **The thermal inertia of near-Earth asteroids**

Editorial correspondence to:

Marco Delbo'

Laboratoire Cassiopée

Observatoire de la Côte d'Azur

BP 4229 06304 Nice, cedex 04 - France

Tel.: +33 (0)4 9200 1944

Fax.: +33 (0)4 9200 3121

E-mail: delbo@obs-nice.fr

*Abstract*

Thermal inertia determines the temperature distribution over the surface of an asteroid and therefore governs the magnitude the Yarkovsky effect. The latter causes gradual drifting of the orbits of km-sized asteroids and plays an important role in the delivery of near-Earth asteroids (NEAs) from the Main Belt and in the dynamical spreading of asteroid families. At present, very little is known about the thermal inertia of asteroids in the km size range. Here we show that the average thermal inertia of a sample of NEAs in the km size range is 200±40 J m$^{-2}$ s$^{-0.5}$ K$^{-1}$. Furthermore, we identify a trend of increasing thermal inertia with decreasing asteroid diameter, $D$. This indicates that the dependence of the drift rate of the orbital semimajor axis on the size of asteroids due to the Yarkovsky effect is a more complex function than the generally adopted $D^{-1}$ dependence, and that the size distribution of objects injected by Yarkovsky-driven orbital mobility into the NEA source regions is less skewed to smaller sizes than generally assumed. We discuss how this fact may help to explain the small difference in the slope of the size distribution of km-sized NEAs and main belt asteroids.

# 1  Introduction

Observations of asteroids in the wavelength range of their thermal-infrared emission (>5 µm) have been used since the 1970s (Allen, 1970) to determine the sizes and the albedos of these bodies. In recent years, thanks to the advances in detector technology and the availability of 10-m class telescopes on the ground, thermal-infrared observations of asteroids have improved in sensitivity. Increased efforts have consequently been devoted to deriving the sizes and albedos of near-Earth asteroids (NEAs; for reviews see Harris and Lagerros, 2002; Delbo' and Harris, 2002; Delbo', 2004; Harris, 2006 and references therein), in order to better assess the impact hazard these bodies pose to our planet and to improve our understanding of their relation to main-belt asteroids and comets (see Stuart and Binzel, 2004; Morbidelli *et al.*, 2002). Furthermore, improvements in spectral coverage and the possibility of easily obtaining spectrophotometric data through narrow-band filters in the range 5 – 20 µm have allowed information on the surface temperatures of asteroids to be obtained. The spectrum of the thermal-infrared radiation received from a body is related to the temperature distribution on that part of its surface visible to the observer. Several factors play a role in determining the temperature distribution on the surface of an asteroid, such as the heliocentric distance, albedo, obliquity of the spin vector, rotation rate, and a number of thermal properties of the surface such as its thermal inertia.

Thermal inertia is a measure of the resistance of a material to temperature change. It is defined as $\Gamma = \sqrt{\rho \kappa c}$, where $\kappa$ is the thermal conductivity, $\rho$ the density and $c$ the specific heat capacity. The thermal inertia of an asteroid depends on regolith particle size and depth, degree of compaction, and exposure of solid rocks and boulders within the top few centimeters of the subsurface (see e.g. Mellon *et al.*, 2000). At the limit of zero thermal inertia (the most simple temperature distribution model for asteroids), a body with a smooth surface would display a temperature distribution which depends only on the solar incidence angle $i$, (on a sphere, $i$ is also the angular distance of a point from the subsolar point):

$$\begin{aligned} T &= T_{SS} \cos^{1/4} i, & i \leq \pi/2 \\ T &= 0, & i > \pi/2 \end{aligned} \quad (1)$$

The subsolar temperature, $T_{SS}$, is determined by equating the total energy absorbed by a surface element to that emitted in the thermal infrared, i.e.:

$$\frac{S_\odot (1-A)}{r^2} = \eta \sigma \varepsilon T_{SS}^4 \quad (2)$$

where $A$ is the bolometric Bond albedo, $S_\odot$ is the solar constant, $r$ is the heliocentric distance of the body, $\varepsilon$ is the infrared emissivity, $\sigma$ is the Stefan-Boltzmann constant and $\eta$ is the so-called "beaming parameter", which is equal to one in the case that each point of the surface is in instantaneous thermal equilibrium with solar radiation. The surface temperature distribution that one obtain for $\eta=1$ on a spherical shape is that of the so-called Standard Thermal Model (STM, Lebofsky and Spencer, 1989) that was widely used to derive diameters and albedos especially of main-belt asteroids (MBAs). In the more realistic case of a body with finite thermal inertia and rotating with a spin vector not pointing toward the sun, the temperature distribution is no longer symmetric with respect to the subsolar point: each surface element behaves like a capacitor or sink for the solar energy such that the body's diurnal temperature profile becomes more smoothed out in longitude (see Spencer *et al.*, 1989; Delbo' and Harris, 2002; Delbo', 2004). The hottest temperatures during the day decrease, whereas those on the night-side do not drop to zero as in the idealistic case of zero thermal inertia, implying non-zero thermal-infrared emission from the dark side of the body.

However, the effect of thermal inertia is coupled with the rotation rate of the body. An asteroid rotating slowly with a high thermal inertia displays a similar temperature distribution to one rotating more rapidly but with a lower thermal inertia. The degree to which the surface of an asteroid can respond to changes in insolation can be characterized by a single parameter: this is the so-called thermal parameter $\Theta$ (e.g. Spencer *et al.*, 1989), which combines rotation period, $P$, thermal inertia, $\Gamma$, and subsolar surface temperature, $T_{SS}$, and consequently depends on the heliocentric distance of the body. The thermal parameter is given by:

$$\Theta = \frac{\Gamma}{\varepsilon \sigma T_{SS}^3} \sqrt{\frac{2\pi}{P}}. \qquad (3)$$

Note that objects with the same value of $\Theta$, although with different $P$ or $\Gamma$ display the same diurnal temperature profile, provided they have the same shape and spin axis obliquity (the angle formed by the object spin vector and the direction to the Sun). In the case of non-zero thermal inertia, because the temperature distribution is no longer symmetric with respect to the direction to the Sun, the momentum carried off by the photons emitted in the thermal infrared has a component along the orbital velocity vector of the body, causing a decrease or increase of the asteroid orbital energy depending on whether the rotation sense of the body is prograde or retrograde. This phenomenon, known as the Yarkovsky effect, (see Bottke *et al.*, 2002) causes a secular variation of the semimajor axis of the orbits of asteroids on a time scale of the order of $10^{-4}$ AU/Myr for a main-belt asteroid at 2.5 AU from the Sun with a diameter of 1 km. The Yarkovsky effect is responsible for the slow but continuous transport of small asteroids and meteoroids from the zone of their formation into chaotic resonance regions that can deliver them to near-Earth space (Bottke *et al.*, 2002; Morbidelli and Vokrouhlický, 2003). The Yarkovsky effect is also important to explain the spreading of asteroid dynamical families (Bottke *et al.*, 2001; Bottke *et al.*, 2006; Nesvorný and Bottke, 2004). Moreover, the emission of thermal photons also produces a net torque that alters the spin vector of small bodies in two ways: it accelerates or decelerates the spin rate and also changes the direction of the spin axis. This mechanism was named by Rubincam (2000) as the Yarkovsky-O'Keefe-Radzievskii-Paddack effect, or YORP for short.

Knowledge of the thermal inertia of asteroids is thus important for a number of reasons: (a) It can be used to infer the presence or absence of loose material on the surface: thermal inertia of fine dust is very low: ~30 J m$^{-2}$ s$^{-0.5}$ K$^{-1}$ (Putzig *et al.*, 2005); lunar regolith, a layer of fragmentary incoherent rocky debris covering the surface of the Moon, also has a low thermal inertia of about 50 J m$^{-2}$ s$^{-0.5}$ K$^{-1}$ (Spencer *et al.*, 1989). Coarse sand has a higher thermal inertia, i.e. about 400 J m$^{-2}$ s$^{-0.5}$ K$^{-1}$ (Mellon *et al.*, 2000; Christiansen *et al.*, 2003), that of bare rock is larger than 2500 J m$^{-2}$ s$^{-0.5}$ K$^{-1}$ (Jakosky, 1986), whereas the thermal inertia of metal rich asteroidal fragments can be larger than 12000 J m$^{-2}$ s$^{-0.5}$ K$^{-1}$ (Farinella *et al.*, 1998, Table 1). (b) Thermal inertia is the key thermophysical parameter that determines the temperature distribution over the surface of an asteroid and therefore governs the magnitude of the Yarkovsky and YORP effects (Capek and Vokrouhlický, 2004). (c) It allows a better determination of systematic errors in diameters and albedos derived using simple thermal models, which make assumptions about the surface temperature distribution and/or neglect the thermal-infrared flux from the non-illuminated fraction of the body (see Spencer *et al.*, 1989, Delbo', 2004, Harris, 2005). However, at present, very little is known about the thermal inertia of asteroids in general, especially in the case of bodies in the km size range.

The thermal inertia of an asteroid can be derived by comparing measurements of its thermal-infrared emission to synthetic fluxes generated by means of a thermophysical model (TPM; Spencer, 1990; Lagerros, 1996; Emery *et al.*, 1998; Delbo', 2004), which is used to calculate the temperature distribution over the body's surface as a function of a number of parameters, including the thermal inertia $\Gamma$. In these models, the asteroid shape is modeled as a mesh of planar facets. The temperature of each facet is determined by numerically solving the one-dimensional heat diffusion equation using assumed values of the thermal inertia, with the boundary condition given by the time-dependent solar energy absorbed at the surface of the facet (see Delbo', 2004). This latter quantity is calculated from the heliocentric distance of the asteroid, the value assumed for the albedo, and the solar incident angle. Macroscopic surface roughness is usually modeled by adding hemispherical section craters of variable opening angle and variable surface density to each facet. Shadowing and multiple reflections of incident solar and thermally emitted radiation inside craters are taken into account as described by Spencer (1990), Emery *et al.* (1998), and Delbo' (2004). Heat conduction is also accounted for within craters (Spencer *et al.*, 1989; Spencer, 1990; Lagerros, 1996, Delbo', 2004). Surface roughness can be adjusted by changing the opening angle of the craters, the density of the crater distribution, or a combination of both. However, Emery *et al.* (1998) have shown that if surface roughness is measured in terms of the mean surface slope, $\bar{\theta}$, according to the parameterization introduced by Hapke (1984), emission spectra are a function of $\bar{\theta}$ only and not of the crater opening angle and crater surface density. We recall here that

$$\tan \bar{\theta} = \int_0^{\pi/2} a(\theta) \tan \theta \, d\theta \qquad (4)$$

where $\theta$ is the angle of a given facet from the horizontal, and $a(\theta)$ is the distribution of surface slopes. The total observable thermal emission is calculated by summing the contributions from each facet visible to the observer. Model parameters (e.g. $\Gamma$, A, $\bar{\theta}$) are adjusted until the best agreement is obtained with the observational data, i.e. the least-squares residual of the fit $\chi^2$ is minimized, thereby constraining the physical properties (albedo, size, macroscopic roughness, and thermal inertia) of the asteroid.

To date, TPMs have been used to derive the thermal inertia of seven large MBAs (Müller, T. G. and Lagerros, 1998; Müller, T. G. and Blommaert, 2004; Mueller, M. *et al.*, 2006b), and five NEAs (Harris *et al.*, 2005; Müller, T. G. *et al.*, 2005; Mueller, M. *et al.*, 2006a, Harris *et al.*, 2007); values derived lie between 5 and ~1000 J m$^{-2}$ s$^{-0.5}$ K$^{-1}$, i.e. $\Gamma$ varies by more than two orders of magnitude. The applicability of TPMs is limited to the few asteroids for which gross shape, rotation period, and spin axis orientation are known. Multi-epoch observations are also required for obtaining a robust estimation of the thermal properties of asteroids via TPM fit.

There is, however, an extensive set of thermal-infrared observations of NEAs in the km size range for which no TPM fit is possible (e.g. Veeder *et al.*, 1989; Harris, 1998; Harris *et al.*, 1998; Harris and Davies, 1999; Delbo' *et al.*, 2003; Delbo', 2004; Wolters *et al.*, 2005). In order to overcome this limitation, we have developed a statistical inversion method, described in Section 2, enabling the determination of the average value of the thermal inertia of NEAs in the km-size range. Our approach is based on the fact that, even though shapes, rotation periods, and spin axis orientations are not known for every NEA, the distribution of these quantities for the whole population can be inferred from published data (La Spina *et al*., 2004; Hahn, 2006).

In Section 3 we compare the result from our statistical inversion method with the values of the thermal inertias of asteroids determined by means of thermophysical models, and we identify a trend of increasing thermal inertia with decreasing asteroid diameter, *D*.

In Section 4 we describe the implications of the trend of increasing thermal inertia with decreasing asteroid diameter, in particular for the size-dependence of the Yarkovsky effect and the size distribution of NEAs and MBAs.

## 2  Determination of the mean thermal inertia of NEAs

The large majority of asteroids for which we have thermal-infrared observations have been observed at a single epoch and/or information about their gross shape and pole orientation is not available, precluding the use of TPMs. In these cases simpler thermal models such as the near-Earth asteroid thermal model (NEATM; Harris, 1998) are used to derive the sizes and the albedos of these objects. The NEATM assumes that the object has a spherical shape, and its surface temperature distribution is described by Eq. (1) and Eq. (2). However, the parameter $\eta$ is not kept constant, as in the case of the STM, but is adjusted in the fitting procedure to allow the model spectral energy distribution to match the observed data. In order to derive a robust estimate of the $\eta$-value the NEATM requires observations at different, ideally well-spaced, wavelengths in the thermal infrared. The parameter $\eta$ can be seen as a measure of the departure of the asteroid temperature distribution from that of the STM and is a strong function of the surface thermal inertia (Spencer *et al*., 1989; Harris, 1998; Delbo', 2004). However, $\eta$ depends also on parameters such as the macroscopic surface roughness, $\bar{\theta}$, the rotation period, $P$, the bolometric Bond albedo, $A$, the thermal-infrared emissivity, $\varepsilon$, the heliocentric distance, $r$, the gross shape of the body, $\mathbb{S}$, the sub-solar latitude, $\theta_{SS}$, the longitude, $\phi_{SE}$, and the latitude, $\theta_{SE}$, of the sub-Earth point (Delbo', 2004). In general we can write that

$$\eta \equiv \eta(\varepsilon, A, r, \Theta(\Gamma, P), \bar{\theta}, \theta_{SS}, \theta_{SE}, \phi_{SE}, \mathbb{S}). \qquad (5)$$

These parameters are usually not known for the individual objects, but their distributions can be estimated (or reasonably assumed) for the entire population. Note that a set of $\theta_{SS}$, $\phi_{SE}$, and $\theta_{SE}$, which depend on the ecliptic longitude $\lambda_0$ and latitude $\beta_0$ of the pole of the body, also defines the value of the solar phase angle, $\alpha$.

Delbo' *et al*. (2003) noted that qualitative information about the average thermal properties of a sample of NEAs could be obtained from the distribution of the $\eta$-values of the sample as function of the phase angle, $\alpha$. In particular, the absence of large $\eta$-values (e.g. $\eta > 2$) at small or moderate phase angles (e.g. $\leq 45^{\circ}$), and the fact that $\eta$ tends to ~ 0.8 for $\alpha$ approaching $0^{\circ}$, was interpreted in terms of the NEAs having low thermal inertias in general. In subsequent work (Delbo', 2004) it was found that for a synthetic population of spherical asteroids with constant values of $A$, $r$, $\Gamma$, $P$, and $\bar{\theta}$, but with pole directions randomly oriented, the distribution of the points in the $(\alpha, \eta)$ plane is strongly dependent on $\Gamma$. By varying $\Gamma$ until the distribution of the synthetic points in the $(\alpha, \eta)$ plane matched the one derived from the observations, Delbo' (2004) obtained a best-fit thermal inertia for the NEAs equal to ~500 J m$^{-2}$ s$^{-0.5}$ K$^{-1}$. Harris (2005), using a similar method on a larger database of η-values and neglecting the effects of surface roughness ($\bar{\theta}$ =0$^o$), derived a best-fit thermal inertia of ~300 J m$^{-2}$ s$^{-0.5}$ K$^{-1}$.

Here we improve on the above-mentioned work by determining the mean thermal inertia of NEAs using a rigorous statistical inversion method, based on the comparison of the distributions of NEATM η-values from the current NEA database vs. α, with that of a synthetic population of asteroids generated through a TPM, using realistic distributions of the input parameters $P$, θ$_{SS}$, θ$_{SE}$, ϕ$_{SE}$, and $A$ derived from the literature (see Table 1 with published η-values from Harris, 1998, Harris *et al.* (1998); Harris and Davies (1999); Delbo' *et al.* (2003); Delbo' (2004); and Wolters *et al.* (2005). La Spina *et al.* (2004) give the distribution of λ$_0$ and β$_0$ for NEAs, and Hahn (2006) that of NEA rotation rates). In the following section we describe our method in detail.

## *2.1 Model parameter space*

As a first step we studied the dependence of η on the relevant parameters of Eq. 5. This was done by choosing typical parameter values and showing how small perturbations of the assumed values affect η. For the purpose of this analysis we assume $A_0$=0.073, $r_0$=1.2 AU (as we will show below, these are the average values of $A$ and $r$ for the NEAs in our sample), $\bar{\theta}_0$=36$^o$ (the value derived for 433 Eros; Domingue *et al.*, 2002), $\mathbb{S}_0$=sphere, and Θ$_0$=1.0. Note that Θ$_0$=1.0 corresponds to a thermal inertia of ~200 J m$^{-2}$ s$^{-0.5}$ K$^{-1}$ for surface temperatures typical of NEAs and $P$ = 6 hours, a rotation period representative of asteroids with sizes between ~0.15 and 10 km (Pravec *et al.*, 2002). We will show in section 2.3 that Γ = 200 J m$^{-2}$ s$^{-0.5}$ K$^{-1}$ is the mean thermal inertia of NEAs. The illumination and observation geometry was varied such that θ$_{SS}$ was uniformly distributed in the range between 0 and π/2 and θ$_{SE}$, ϕ$_{SE}$ were varied in such a manner that the resulting sub-Earth vectors were uniformly distributed over the celestial sphere. The values of θ$_{SE}$, ϕ$_{SE}$ were further subject to the constraint that the phase angle be ≤ 100$^o$. Figure 1 shows the sensitivity of η to a change in the model parameters. In particular, for each value of θ$_{SS}$, θ$_{SE}$, and ϕ$_{SE}$, the variation of η due to a 1% change in each parameter is plotted. We have also calculated, for some fixed illumination and observation geometries (e.g. θ$_{SE}$ = 0$^o$, ϕ$_{SE}$ = 45$^o$ and θ$_{SS}$ = 0$^o$), how the variations Δη scale with changes in the model parameters. We found that Δη is proportional to Δ$A$, Δ$\bar{\theta}_0$, ΔΘ, and Δ$r$ within a large range of variation (>100%) of each parameter from its nominal value. Because for common asteroidal material the thermal-infrared emissivity is thought to be relatively constant, it has been fixed for this study at ε = 0.9. It is appropriate for objects with surfaces that emit a substantial portion of their thermal-infrared radiation shortward of 8 μm (Lim *et al.*, 2005). Mustard and Hays (1997) have also shown that the reflectance spectra of fine-grained particulate materials, thought to be representative of planetary regoliths, have values around 0.1 and in general smaller than 0.2 in the region 8 – 24 μm. Because the reflectance, $R$, and the emissivity are related by Kirchhoff's law ($R$=1-ε), the measurements cited above implies that ε = 0.9 is a reasonable estimate for the thermal-infrared emissivity of NEA surfaces. Moreover, from Eq. 2 one can calculate that Δη≈1.6Δε for ε≈0.9 and η≈1.5 (the average η for the NEAs for which this parameter was derived from observations; see Table 1). This implies that variations of ε in the range 0.8 – 1.0 cause changes of η that are within the typical uncertainty of ~20% in the estimation of η from observations. Note that the value of Δη/η = 20%, where Δη is the uncertainty in η, is based on the reproducibility of η for those objects for which more than one measurement is available from independent data sets. Moreover, 20% is also the mean value of Δη/η of the "Delbo' Thermal Infrared Asteroid Diameters and Albedos" database at the NASA PDS (Delbo', 2006). In this dataset for those observations where Δη is present, its value was formally calculated from the measurements of the asteroids' thermal infrared fluxes.

The vast majority of the observations in our sample was obtained at a phase angle smaller than 80°, and within this range, Fig. 1 shows that the largest variation of η caused by a 1% change of $A$ (the bolometric Bond albedo) is approximately 0.1%. Because the mean value of $A$ for our sample is $\langle A \rangle = 0.073$ and the standard deviation is 0.04 (see Table 1), the variation of η due to the distribution of the albedos is smaller than 5% and thus small compared to the typical uncertainty of Δη/η ~ 20%. For this reason we have utilized a constant value of 0.073 for $A$ in our statistical inversion method.

Moreover, the variation of η due to a 1% change in the macroscopic surface roughness is strongly phase angle dependent, but in general smaller than 0.2% for α in the range 0-60°. This implies that even a ±100% change in $\bar{\theta}$ causes a variation of η within the typical 20% uncertainty. Note that a ±100% change in $\bar{\theta}$ corresponds to a large variation of the macroscopic roughness, ranging from that of a completely smooth surface to one oversaturated by hemispherical craters. For those observations carried out at α > 60°, η is more sensitive to variations of $\bar{\theta}$. For the reasons above we have treated $\bar{\theta}$ as a free parameter in the inversion method and searched for the value that best fits the observational data.

The sensitivity of η to changes of the objects' heliocentric distances is such that a 1% change of $r$ corresponds to a maximum 0.7% change of η. As calculated for the values in Table 1, the heliocentric distances in our sample have a mean value of 1.2 AU and a standard deviation of 0.1 AU (~8%). The corresponding variation of η is approximately 6% and therefore small. We thus took a constant value of 1.2 AU for $r$ in our statistical inversion method. Only in two cases, namely those of the 29-06-1998 observation of (433) Eros and for the 22-03-2002 observation of (6455) 1992 HE is the variation of η due to the deviation of the heliocentric distance from the nominal value of 1.2 AU slightly larger than the error bars.

We note here that Eq. (5) implicitly contains the assumption that seasonal effects do not affect asteroid surface temperatures. However, when $\Theta \neq 0$, asteroid temperatures always depend on the previous thermal history of the surface. Since NEAs have in general large orbital eccentricities, these bodies experience large variations of insolation as a function of their orbital position, which may lead to a seasonal component of the variation of their surface temperatures and thus of the corresponding η-values. To demonstrate that our working hypothesis of Eq. (5) is valid (i.e. seasonal components are negligible), we calculated η for several synthetic asteroids with the same physical characteristics, but with different orbits, in order to explore the effect of different levels of insolation. Orbits were chosen with eccentricities in the range 0 to 0.8 but with a common perihelion distance $r_p$. For different values of the asteroid thermal inertia in the range 200-5000 J m$^{-2}$ s$^{-0.5}$ K$^{-1}$ and $r_p$ in the range 0.5 - 1.5 AU, we found variations of only a few percent in the η-values calculated at $r_p$. This leads us to conclude that seasonal variations in the η-values are small and that Eq. (5) is valid.

In general NEAs have elongated shapes, which may cause their surface temperature distributions to differ from that of a spherical object with the same surface properties and illumination geometry. We studied the sensitivity of η to deviations from the spherical shape by calculating η-values of tri-axial ellipsoids, the semiaxes of which were varied in the ratio $a/(\sqrt{a}/1)$ with $1 \leq a \leq 6$, assuming $A_0$=0.073, $r_0$=1.2 AU, $\bar{\theta}_0$=36°, $\Theta_0$=1.0, and for random orientations of the shape with respect to the Sun and the Earth. We found that the distribution of Δη is a function of $a$ (with values of Δη increasing with increasing $a$), where Δη is the deviation of η from that calculated using a sphere under the same illumination and viewing geometry. However, the relative error on η, Δη/η, is always smaller than ±10% for $a \leq 5$ and $\alpha \leq 45°$. For $\alpha > 45°$, the mean value of the relative error, ⟨Δη/η⟩, is smaller than +15% and its standard deviation, $\sigma_{\Delta\eta/\eta}$, is smaller than 5% for $a \leq 5$. Because the maximum lightcurve amplitude of our model ellipsoid is $L\approx1.25\log a$ mag, Δη/η is smaller than 20% if $L \leq 0.873$ mag. This condition is in general satisfied for the NEAs in Table 1, for which the average value of L is around 0.6 mag.

We expect that the contributions to Δη due to variations of the model parameters $A$, $r$, $\bar{\theta}$, and the ellipsoid axial ratio $a$, stack up randomly, since deviations of these parameters from their mean values are fully uncorrelated (e.g. there is no apparent reason that an NEA with an albedo higher than the average is also observed at an heliocentric distance higher than the its average value). We performed some numerical experiments in order to cross check this assumption and found that the value of Δη is in general a good proxy of $[(\partial\eta/\partial A\Delta A)^2 + (\partial\eta/\partial r\Delta r)^2 (\partial\eta/\partial\bar{\theta}\Delta\bar{\theta})]^{1/2}$. Adding the effect of non-spherical shapes increases the value of Δη, but never systematically at phase angles < ~60°. It is clear that ellipsoids are highly idealized shapes and larger contributions to Δη may be expected in the case of real NEAs.

Figure 1 shows that the sensitivity of η to changes in the thermal parameter is very similar to the sensitivity to changes in $r$, with variations of η in general no larger than 0.5% for a 1% change of Θ. However, while the value of $r$ for the asteroids in Table 1 is rather constant around the mean value of 1.2 AU, Θ can range between 0.1 and 20 considering that thermal inertia can be anywhere between 10 J m$^{-2}$ s$^{-0.5}$ K$^{-1}$ (the thermal inertia of large main-belt asteroids) and 2500 J m$^{-2}$ s$^{-0.5}$ K$^{-1}$ (that of bare rock). This implies that the scatter in the η-values that we observe in the NEAs of Table 1 is mainly a function of α and Θ, which depends on the thermal inertia. If we assume the thermal inertia to be roughly constant within the NEA population for a given size, its value can be inferred from the distribution of the measured η-values versus α. This is the idea on which our statistical inversion method is based.

## *2.2 Model populations*

Our inversion method requires η to be computed for all members of a synthetic population of NEAs as a function of Γ. The calculation of η was performed by numerically generating thermal-infrared spectra by means of a TPM and fitting them with the NEATM. As discussed in the previous sections, the parameters $A$, ε, r, $\bar{\theta}$, and $\mathbb{S}$ contribute little to the variation of η within the expected parameter ranges. Therefore, they have been kept fixed to their nominal values throughout the modeling process. In order to keep the amount of computing time required for the inversion within reasonable limits, the values of η have been computed only once for all possible combinations of the remaining parameters, and the results have been stored in a four-dimensional look up table. The granularity of the look up table was chosen to be small enough to cause changes of η of about 0.1 between two consecutive parameter steps.

For each value of Γ, we then generated a large number (30,000) of synthetic objects whose parameters have random values with distributions that have been chosen to provide a reasonable match to the observed population of NEAs. In particular:

(i) the distributions of the angles $\theta_{SS}$, $\theta_{SE}$, and $\phi_{SE}$ were computed starting from the distributions of the spin-axis orientation ($\lambda_0$, $\beta_0$) from La Spina *et al*. (2004), the phase angle α, the heliocentric ecliptic latitude $\beta_H$, and the geocentric ecliptic latitude $\beta_E$ of the asteroids at the time of the infrared observations (see Table 1 and Fig. 2);

(ii) the distribution of the thermal parameter was calculated starting from the distribution of the NEA rotation periods (Hahn, 2006) and by using a constant value of Γ.

In Fig. 3 three such populations are shown that correspond to the Γ values of 15 (green), 200 (red), and 1000 (blue) J m$^{-2}$ s$^{-0.5}$ K$^{-1}$, respectively. We have superimposed the η values for the NEAs in Table 1 on the synthetic data plot.

## *2.3 Best-fit procedure*

Figure 3 gives a clear visual impression of the dependence of η on Γ. We therefore used a formal best-fit technique to estimate the value of Γ for which a synthetic population best fits the observed data, under the assumption that Γ is constant for all objects in the observed sample. The method that we used to compare the observed data with the bi-dimensional distributions of the synthetic points in the (α, η) plane is based on the two-dimensional Kolmogorov-Smirnov metric (K-S; Press *et al*., 1992). The distance *D* of the K-S metric is used as the goodness of fit estimator (Press *et al*., 1992). Our best-fit procedure consisted of finding the value of Γ that minimizes the K-S distance *D*. From here on, we indicate this value with the symbol Γ$^*$. Figure 4, where we have plotted the K-S distance *D* as a function of Γ, shows that the function *D* (Γ) has a minimum at Γ=200 J m$^{-2}$ s$^{-0.5}$ K$^{-1}$, which is the value of thermal inertia that we take for Γ$^*$.

We expect Γ$^*$ to depend upon the assumed value for $\bar{\theta}$, the distributions of NEA rotation rates, and also on the spin-axis orientations that we have used to produce the distribution of the input parameters $\theta_{SS}$, $\theta_{SE}$, and $\phi_{SE}$. Moreover, the value of Γ$^*$ must be affected by the errors in the measurements of the thermal infrared fluxes, i.e. by the errors on the η-values taken from the literature. In order to study the sensitivity of Γ$^*$ to changes applied to the nominal values of the input parameters, we first varied $\bar{\theta}$ in the range between 0° (perfectly smooth surface) and 58° (corresponding to the surface completely covered by hemispherical craters). Figure 4 shows the function *D* (Γ) for three different values of $\bar{\theta}$. It clearly demonstrates that the value of Γ$^*$ only weakly depends on $\bar{\theta}$ and that a high degree of surface roughness produces a better fit to the observed data.

We also investigated the sensitivity of Γ$^*$ to changes in the input distributions of asteroids' spin-axis orientations and rotation rates. Figure 4 shows the function *D* (Γ) obtained by using random spin-axis orientations uniformly distributed over the sphere instead of the nominal distribution. In that case, the best-fit thermal inertia increases to 250 J m$^{-2}$ s$^{-0.5}$ K$^{-1}$, and to 230 J m$^{-2}$ s$^{-0.5}$ K$^{-1}$ if the distribution of the rotation rates are assumed to be uniformly distributed between 4 and 10 hours, a case which we believe to be very extreme.

The sensitivity of $\Gamma^*$ to the errors affecting the η-values from Table 1 was studied by performing extended Monte Carlo simulations, in which we randomly varied the values of the η-values within their error bars (using normally-distributed random numbers), and for each simulation of noise-corrupted data we calculated the best-fit thermal inertia. The standard deviation of $\Gamma^*$ was found to be 40 J m$^{-2}$ s$^{-0.5}$ K$^{-1}$.

Of course, we expect that the distribution of the data points in Fig. 3 derives from a population with a range of thermal inertias, and further investigation is required to understand what the relations are between $\Gamma^*$ and the parameters defining the population, such as the mean value of Γ and the standard deviation of its distribution. In order to answer this question, we applied our inversion method on (α, η) points obtained from synthetic populations of NEAs with known distributions of thermal inertia. We used random values of Γ uniformly and normally distributed, varying both the mean value and the standard deviation of the populations. We found that our fitting procedure, based on the minimization of the K-S distance *D*, is capable of retrieving a good estimate of the mean value for Γ of the populations in all cases. We conclude that the average value of the thermal inertia for km-sized NEAs is 200±40 J m$^{-2}$ s$^{-0.5}$ K$^{-1}$, which is about four times that of the lunar soil and corresponds to a surface thermal conductivity of $0.027^{+0.015}_{-0.010}$ W m$^{-1}$ K$^{-1}$ assuming that the surface material density and specific heat capacity are in the range 1500-3500 kg m$^{-3}$ and 500-680 J kg$^{-1}$ K$^{-1}$, respectively (Britt *et al*., 2002; Farinella *et al*., 1998).

The value of $\Gamma^*$ that we have derived by means of the best fit procedure is less than 10% of that expected for a bare-rock surface (Jakosky, 1986). This implies that the surfaces of NEAs have in general significant quantities of thermally-insulating regolith. However, $\Gamma^*$ is also about four times higher than the value that has been determined for the lunar soil and more than ten times higher than the thermal inertia typical of large main-belt asteroids. This effect may be due to the fact that the regolith present on NEA surfaces is less mature and/or less thick than that of the Moon and the largest MBAs. The higher NEA thermal inertia can also be explained in terms of a coarser regolith and the exposure of rocks and boulders on the surface of these bodies, as clearly shown in the high resolution images of (433) Eros and (25143) Itokawa obtained by the NEAR Shoemaker and the Hayabusa missions, respectively.

A population of asteroids with constant Γ=200 J m$^{-2}$ s$^{-0.5}$ K$^{-1}$ gives the best fit to the dataset. Figure 3 shows, however, that five points with η > 2 are clearly significantly higher than the majority, indicating that these objects presented unusually low color temperatures to the observer, possibly due to higher-than-average thermal inertia (see Delbo' *et al*., 2003 and Delbo', 2004). To gain insights into the *width* of the distribution of the thermal inertia of km-sized NEAs, we fitted the observed distribution of the data points with a synthetic population in which Γ was assumed to be uniformly distributed between 0 and $\Gamma_{MAX}$. The best fit was obtained for $\Gamma_{MAX}$ ~ 600 J m$^{-2}$ s$^{-0.5}$ K$^{-1}$. This suggests that the large majority of km-sized NEAs in our sample have thermal inertia below this value.

The average value of the thermal inertia was derived for a sample of objects whose diameter distribution is shown in Fig. 5. We use here the radiometric diameters as derived by the NEATM. The mean diameter of the sample is 3 km, but if we remove the asteroid 433 Eros, the mean diameter value decreases to 2 km. 433 Eros is much larger than the average size of the sample (see Fig. 5). In fact the median value of the diameter distribution (including 433 Eros) is 1.8 km. We note that the distribution of log *D* (where *D* is the diameter measured in km) is well fitted by a Gaussian distribution with a central value of 1.7 km. The standard deviation of the best-fit Gaussian function is 0.31 (in log *D*). We can thus conclude that the average value of the thermal inertia is representative of NEAs in the size range 0.8 – 3.4 km.

## 3  Size dependence of asteroid thermal inertia

The mean thermal inertia for the sample of NEAs with published η-values is consistent with the values derived by means of TPMs for (433) Eros (Mueller, M. *et al.*, 2006a), (1580) Betulia (Harris *et al.*, 2005), (25143) Itokawa (Mueller, M. *et al.*, 2006a; Müller, T. G. *et al.*, 2005), and (33342) 1998 WT$_{24}$ (Harris *et al.*, 2007) for which values around 150, 180, 350, 630, and 200 J m$^{-2}$ s$^{-0.5}$ K$^{-1}$ have been obtained respectively. Note that in the case of (25143) Itokawa, Müller, T. G. *et al.* (2005) have obtained a thermal inertia value of 750 J m$^{-2}$ s$^{-0.5}$ K$^{-1}$ combining thermal-infrared observations gathered at ESO in 2004 with those obtained by Delbo' (2004) in 2001. On the other hand, from the latter dataset of observations and a series of further observations of (25143) Itokawa obtained at the NASA-IRTF 3 m telescope with MIRSI in 2004, Mueller, M. *et al.* (2006a) derived a thermal inertia of ~350 J m$^{-2}$ s$^{-0.5}$ K$^{-1}$ or ~800 J m$^{-2}$ s$^{-0.5}$ K$^{-1}$ depending on whether the size of the body was obtained from the TPM or was forced to the radar value of Ostro *et al.* (2004). In this work we have taken the mean value and the extreme values of 350, 750, and 800 J m$^{-2}$ s$^{-0.5}$ K$^{-1}$ as our best estimate for the thermal inertia of Itokawa and its uncertainty. Müller T. G. *et al.* (2004) have also attempted at deriving the thermal inertia of the small (~0.28 km) NEA 2002 NY$_{40}$. They obtained a value of 100 J m$^{-2}$ s$^{-0.5}$ K$^{-1}$ in the case that the size of the object was derived from the TPM, or 1000 J m$^{-2}$ s$^{-0.5}$ K$^{-1}$ if the body's size was forced to the value obtained from radar observations. However, it is important to note that that the thermal inertia of 2002 NY$_{40}$ was derived by assuming an equator-on view and a spherical shape for this object. The value of the thermal inertia derived from the TPM is in general strongly dependent on the pole orientation of the body. For this reason we expect the value of Γ derived for 2002 NY$_{40}$ be less reliable than the values obtained for the other NEAs, for which the pole orientation derived from lightcurve inversion was adopted.

From thermophysical modeling, Müller, T.G. and Lagerros (1998) derived the thermal inertias of a number of the largest MBAs, namely (1) Ceres, (2) Pallas, (3) Juno, (4) Vesta, and (532) Herculina, obtaining the values of 10, 10, 5, 25, and 15 Jm$^{-2}$ s$^{-0.5}$ K$^{-1}$, respectively. Using the same approach, Müller, T.G. and Blommaert (2004) derived a thermal inertia of 15 J m$^{-2}$ s$^{-0.5}$ K$^{-1}$ for (65) Cybele, and Mueller, M. *et al.* (2006b) obtained Γ~50 J m$^{-2}$ s$^{-0.5}$ K$^{-1}$ for (21) Lutetia. From the published plots of the goodness of the TPM fit to the thermal-infrared data as a function of Γ it is possible to deduce that the relative uncertainties for the thermal inertias of these asteroids are around 50%.

From the comparison of the values of Γ mentioned above, it is clear that there is an increase in the thermal inertia from that of large MBAs with diameters of several hundred km to that of much smaller km-sized NEAs, and that the values of Γ obtained for km-sized NEAs are about one order of magnitude or more higher than the values derived for large MBAs, but still an order of magnitude lower than the thermal inertia of bare rock (~2500 J m$^{-2}$ s$^{-0.5}$ K$^{-1}$; Jakosky, 1986). In order to highlight the behavior of the thermal inertia of asteroids as a function of their size, we have plotted the mean value of thermal inertia for NEAs and the values of the thermal inertia derived by means of TPMs against object diameters in Fig. 6. Small open circles represent the literature values derived from the application of TPMs. The large open diamond is the result from this work. The axis on the right-hand side gives the asteroid surface thermal conductivity $k$ as a function of size, on the basis of $k=\Gamma^2/(\rho c)$, with constant surface density $\rho = 2500$ kg m$^{-3}$ and specific heat capacity $c = 600$ J kg$^{-1}$ K$^{-1}$. These values are reasonable assumptions for asteroid surfaces (Britt *et al.*, 2002; Farinella *et al.*, 1998). For the asteroid 2002 NY$_{40}$ a bar between 100 and 1000 J m$^{-2}$ s$^{-0.5}$ K$^{-1}$ is drawn. The thermal conductivity has also been constrained in the cases of (6489) Golevka (Chesley *et al.*, 2003) and for asteroids in the Karin cluster (Nesvorný and Bottke, 2004). The values of the thermal conductivities derived by these authors have been converted to values of $\Gamma$ assuming $\rho=2500$ kg m$^{-3}$ and $c=600$ J kg$^{-1}$ K$^{-1}$. Fig. 6 shows that the resulting limits, based on the measurements of the Yarkovsky effect on these bodies, are in general agreement with our results.

Figure 6 reveals a convincing trend of increasing thermal inertia with decreasing asteroid diameter, *D*, confirming the intuitive view that large main-belt asteroids, over many hundreds of millions of years, have developed substantial insulating regolith layers, responsible for the low values of their surface thermal inertia. On the other hand, much smaller bodies, with shorter collisional lifetimes, presumably have less regolith, or less mature regolith, and therefore display a larger thermal inertia. Deriving a functional dependence of the thermal inertia as a function of the size of the body has important implications for improving the models of the orbital mobility of asteroids due to the Yarkovsky effect and to better quantify systematic errors in radiometric diameters and albedos of small bodies based on the use of thermal models that neglect the effects of heat conduction, such as the STM. The graph in Fig. 6 suggests that, to the first order, thermal inertia in this size range follows a power law. Expressing $\Gamma$ as

$$\Gamma = d_0 D^{-\xi} \qquad (6)$$

(a linear relation in the log $\Gamma$ – log D plot), a linear regression gives best-fit values of $\xi=0.48\pm0.04$ and $d_0=300\pm47$, where *D* is km and $\Gamma$ in S. I. units (J m$^{-2}$ s$^{-0.5}$ K$^{-1}$), and the 1$\sigma$ uncertainty is based on the assumption that the errors on the thermal inertia and diameter values are normally distributed. (The values of $\Gamma$ for 2002 NY$_{40}$, 6489 Golevka and the Karin cluster asteroids were excluded from the linear regression analysis).

However, the slope $\xi$ of Eq. (6) may assume different values in different size ranges, since there are reasons to suspect that the surface properties of large asteroids may be different to those of smaller bodies: for example, Bottke *et al.* (2005) showed that asteroids with $D > 100$ km and most bodies with $D > 50$ km in size are likely to be primordial objects that have not suffered collisional disruption in the past 4 Gy. These large bodies have spent sufficient time in the asteroid belt to build a regolith such that they would display a low thermal inertia independent of size. In this case $\xi$ should be about zero for *D* larger than about 50 km. In the same study it was shown that asteroids smaller than ~30 km are statistically the remnants of catastrophic collisional disruption of larger parent bodies, and the smaller the object, the fresher the surface. In this latter case one may intuitively expect that a dependence of the thermal inertia on the asteroid diameter would be more likely to occur, implying $\xi > 0$ for D < 30 km. For these reasons we tried to fit the data piecewise, separating the NEAs from the MBAs: a linear regression of Eq. (6) for the MBAs only of Fig. 6, gives best-fit values of $\xi=0.49\pm0.27$ and $d_0=300\pm150$ (Fig. 6, dotted-line) in good agreement with the trend obtained by fitting the whole dataset of thermal inertias. However, we note that the accuracy of this fit is poor and that the value of ξ is strongly influenced by the thermal inertia of 21 Lutetia. On the other hand, a fit of Eq. (6) for near-Earth asteroids only, gives best-fit values of ξ=0.36±0.09 and $d_0$=300±45 (Fig. 6, dashed-line) which corresponds to a shallower dependence of Γ on D for sizes up to 20 km.

A further distinction in the thermal properties of MBAs compared to that of NEAs is given by the different mean heliocentric distances of the two classes of body, causing NEAs to have average temperatures ~200 K higher than those of MBAs. The thermal conductivity in the regolith is temperature dependent (Keihm, 1984), and so is thermal inertia. This temperature dependence of Γ may alter the slope ξ of Eq. (6) when both NEAs and main-belt asteroids are included in the fit. Under the assumption that heat is transported in the regolith mainly by radiative conduction between grains, the thermal conductivity is proportional to $T^3$, with T being the temperature of the regolith grains (Kührt and Giese, 1989; Jakosky 1986). In this case $\Gamma \propto T^{3/2}$ and, from Eq. (2), $\Gamma \propto r^{-3/4}$, where r is the heliocentric distance of the body. On the basis of this dependence of Γ with respect to r, we corrected the values of the thermal inertias of the asteroids of Fig. 6 to the mean heliocentric distance $r_{ref}$ of 1.7 AU. Although the correction factors are in general smaller than the errors affecting the values of Γ, the thermal inertia values of NEAs ($r < r_{ref}$) are reduced, whereas those of MBAs ($r > r_{ref}$) are increased, yielding a smaller value of the slope ξ=0.37±0.04 and $d_0$=230±30.

Furthermore, the make up of NEA surfaces can be modified by processes such as close encounters with planets causing tidal disruption that do not affect asteroids in the Main Belt. Such processes might have been able to alter or strip off the regolith of some NEAs. Thus, while NEAs may be a good proxy for small main-belt asteroids, more observations are needed to confirm this point.

It is clear that with the present small number of asteroids for which we have an estimate of the thermal inertia it is difficult to reveal possible variations of ξ with respect to the mean trend, ξ ~0.4, in different size ranges. Nevertheless, Fig. 6 shows a clear correlation of Γ with asteroid size and that asteroids in the 1 – 30 km size range have values of Γ in general larger than 100 J m$^{-2}$ s$^{-0.5}$ K$^{-1}$. The fact that thermal inertia increases with decreasing size and that the value of Γ for km and multi-km sized asteroids is at least ten times larger than the value derived for the largest main-belt asteroids, has a number of important implications. First of all, radiometric diameters and albedos of asteroids derived by means of thermal models neglecting the effects of thermal inertia, such as the STM, are likely to be affected by increasing systematic errors with decreasing size. Spencer *et al.* (1989) have studied systematic biases in radiometric diameter determinations as a result of the effects of thermal inertia, rotation rate, pole orientation, and temperature. They concluded that the STM systematically underestimates the diameters of objects with non-negligible thermal inertia, while overestimating their albedos. Because we find that thermal inertia increases with decreasing asteroid diameter, it is likely that the systematic underestimation of asteroid diameters (and overestimation of asteroid albedos) obtained from the STM increases for decreasing asteroid size.

Moreover, the absolute value and size dependence of the thermal inertia for asteroids with diameters smaller than about 10-20 km have crucial implications for the magnitude of the Yarkovsky effect, which is an important phenomenon that offers an explanation for the dispersion of asteroid dynamical families and the slow but steady injection of bodies into the dynamical resonances that eventually transport them from the main belt to near-Earth space.

## 4 Implications for the magnitude of the Yarkovsky effect

Current models of Yarkovsky-assisted delivery of NEAs from the Main Belt (Morbidelli and Vokrouhlický, 2003) and the spreading of asteroids families (Bottke *et al.*, 2001; Nesvorný and Bottke, 2004), assume that thermal inertia is independent of object size. In this case, the theory of the Yarkovsky effect predicts that the orbital semimajor axis drift rate of an asteroid, $da/dt$, is proportional to $D^{-1}$ (Bottke *et al.*, 2002). However, the mean value of the thermal inertia derived for NEAs and the inverse correlation of this thermophysical property with asteroid size, demonstrated in this work, give rise to a different magnitude of the Yarkovsky effect and a modified dependence of $da/dt$ on the object diameter. In order to derive this modification one can directly insert the function $\Gamma = d_0 D^{-\xi}$ in the formulas given by Bottke *et al.* (2002) and Vokrouhlický (1999) to explicitly calculate $da/dt$ as a function of the relevant parameters.

Here we discuss the case of MBAs with $D < 10$ km, assuming the linearized theory of the diurnal component of the Yarkovsky effect of Vokrouhlický (1999), which yields

$$\frac{da}{dt} \propto \frac{1}{D} \frac{\Theta}{1+\Theta+0.5\Theta^2}. \tag{7}$$

Because $\Theta$ is directly proportional to $\Gamma$, for $\Theta \gg 1$, $da/dt \propto D^{-1}\Gamma^{-1}$ and hence $da/dt \propto D^{\xi-1}$. We found that this condition holds in general for small asteroids in the Main Belt: in fact for objects with $D$ smaller than ~10 km Fig. 6 shows that the thermal inertia is in general >100 Jm$^{-2}$s$^{-0.5}$K$^{-1}$, and at heliocentric distances >2 AU the surface temperatures of these bodies are in general smaller than 250 K (see e.g. Delbo', 2004). Pravec *et al.* (2002) have also shown that asteroids with sizes between ~0.15 and 10 km have a typical rotation rate around 6 hours. Inserting these values in Eq. (3), we find that $\Theta$ is in general larger than ~2 for main-belt asteroids in this size range. By taking $\xi \sim 0.4$, as derived in the previous section for the $\Gamma = d_0 D^{-\xi}$ relation, we obtain that the size dependence of $da/dt$ due to the diurnal component of the Yarkovsky effect is proportional to $D^{-0.6}$ rather than proportional to $D^{-1}$, as generally assumed, which would hold true for a thermal inertia independent of asteroid size.

We caution, however, that there are currently no reliable estimates of thermal inertia available for any object smaller than Itokawa (D ~ 350 m), so the relation derived for $da/dt \propto D^{-0.6}$ should be assumed to hold for objects with diameters in the range between ~0.35 and ~10 km. Moreover, for asteroids smaller than 350 m and/or higher values of the thermal inertia, the seasonal component of the Yarkovsky effect may become significant and contribute to the average value of *da/dt*. From this analysis we conclude that, in the Main Belt, the drift rate in semimajor axis due to the diurnal component of the Yarkovsky effect increases with decreasing asteroid size more slowly than is normally assumed in models of the origin of NEAs and the spreading of asteroids families.

The shallower dependence of the Yarkovsky effect on the diameter of the bodies caused by the inverse correlation of $\Gamma$ with $D$ has the important implication that the size distribution of the asteroids injected into the NEA source regions is less skewed to small objects than generally assumed. In the following we briefly discuss some of the consequences of this: there is general consensus that the large majority of NEAs originate from the Main Belt via well defined "feeding zones" of dynamical instability (Morbidelli *et al.*, 2002). Asteroidal material can gradually drift towards these NEA source regions as a result of Yarkovsky-driven semimajor axis mobility (Morbidelli *et al.*, 2002; Morbidelli and Vokrouhlický, 2003). The cumulative size distribution of a population of asteroids in a given diameter range (e.g. 0.35 < D < 10 km) can be approximated by a simple exponential function of the form

$$N(>D) = N_0 D^{-\alpha}, \qquad (8)$$

Therefore, according to this asteroid delivery model, the difference in the exponent α between the bodies injected into the NEA source regions and the remaining population of asteroids in the Main Belt is of the order of ~1 if the semimajor axis mobility is proportional to $D^{-1}$. The same difference in the value of the exponents holds for the NEA and the MBA populations in a comparable size range, assuming that the large majority of NEAs come from the Main Belt.

The results of the latest studies that have analyzed the size distributions of NEAs and km-sized MBAs imply that this difference is closer to 0.5-0.7, in favor of a Yarkovsky effect less effective for smaller asteroids, which is consistent with the results of this work. We recall that Eq. (8) can be converted into a cumulative absolute visual magnitude $H$ distribution of a population of asteroids with the form

$$N(<H) = N_0' 10^{\beta H}, \qquad (9)$$

where the exponential slope of the absolute magnitude distribution, β, can be converted into the power-law slope of the diameter distribution via α = 5β (see, e.g., Stuart and Binzel, 2004) . Several authors (Rabinowitz *et al.*, 2000; Bottke *et al.*, 2000; Stuart and Binzel, 2004) agree that β is in the range 0.35 – 0.39 for the NEA population, which implies a value of $\alpha_{NEA}$ between 1.75 and 1.95. The size distribution of km- and sub km-sized MBAs is less constrained than that of NEAs, since the known population is still rather incomplete for $H > 14 - 15$ (corresponding to values of $D$ between 6 and 3 km for a geometric visible albedo of 0.11), so that beyond this threshold only extrapolations of the known distribution can be made. Morbidelli and Vokrouhlický (2003) used the slopes derived by Ivezic *et al.* (2001) from the Sloan Digital Sky Survey (SDSS), namely, β=0.61 for 13 < *H* < 15.5 and β=0.25 for 15.5 < *H* < 18, to extrapolate the observed *H* cumulative distribution (as given by the Astorb catalog) to km sized asteroids, and use it in their model of the Yarkovsky-driven origin of near-Earth asteroids. Assuming β=0.25 for 15.5 < *H* < 18 this would give a value of $\alpha_{MBA}$=1.25 for the slope of the cumulative size distribution of MBAs in the range 1 < *D* < 3 km. This value is in good agreement with the even slightly shallower size distribution ($\alpha_{MBA}$~1.2) of km and sub km sized MBAs (for 0.5 < *D* < 1 km) found by the SMBAS survey (Sub-km Main-Belt Asteroid Survey) obtained by Yoshida *et al.*, (2003). Taking the values of α from the studies above, we find that $\alpha_{NEA} - \alpha_{MBA} = 0.5 - 0.7$, in good agreement with a drifting population of asteroids with $da/dt \propto D^{-0.6}$.

However, Morbidelli and Vokrouhlický (2003) have also shown that the collisional re-orientation of asteroid spin axes (which resets the drift speed due to the Yarkovsky effect), the collisional disruption of the bodies during their slow drift towards the NEA source regions, and the YORP effect, tend to decrease the difference between $\alpha_{NEA}$ and $\alpha_{MBA}$. The addition of these phenomena along with the revised dependence of $da/dt \propto D^{-0.6}$ due to the Yarkovsky effect may help to explain the even steeper size distribution of small MBAs, and thus a difference between $\alpha_{NEA}$ and $\alpha_{MBA}$ smaller than 0.5 – 0.7 implied by the recent results of the Sub-Kilometer Asteroid Diameter Survey (SKADS; Davis *et al*., 2006), which found $\beta_{MBA}$ = 0.38, corresponding to $\alpha_{MBA}$ = 1.9, for 13 < *H* < 17.

The value of the thermal inertia also plays an important role in the YORP effect (Rubincam, 2000; Vokrouhlický and Capek, 2002), which is a torque produced by the thermal radiation emitted by asteroids with irregular shapes causing a slow spin-up/spin-down and a change of the spin axis obliquities of these bodies. In contrast to the Yarkovsky effect, YORP also acts on bodies with zero surface thermal conductivity. However, in the case of a thermal inertia significantly larger than zero (in contrast to the case of zero-conductivity), YORP preferentially drives obliquity toward two asymptotic states perpendicular to the orbital plane, and asymptotically decelerates and accelerates rotation rate in about an equal number of cases (Capek and Vokrouhlický, 2005). Capek and Vokrouhlický (2005) have shown that the acceleration of the rotation rate, $d\omega/dt$, is largely independent of the thermal inertia, whereas its value significantly affects the rate of change of the obliquity, $d\theta_{SS}/dt$, in the sense that the higher the thermal inertia the larger the mean value of $d\theta_{SS}/dt$. Capek and Vokrouhlický (2005) found the median value of the distribution of $d\theta_{SS}/dt$ for populations of Gaussian spheres increased from 3.33 deg/My for $\Gamma$=0 J m$^{-2}$ s$^{-0.5}$ K$^{-1}$ to 5.94 deg/My for $\Gamma$=39 J m$^{-2}$ s$^{-0.5}$ K$^{-1}$ and to 8.60 deg/My in the case of $\Gamma$=122 J m$^{-2}$ s$^{-0.5}$ K$^{-1}$, which is a suitable value for bodies of about 5 km according to our Fig. 6. Because our $\Gamma$ (*D*) relation predicts an even larger value of surface thermal inertia for asteroids of 1 km in diameter, the YORP reorientation of the spin vector of asteroids becomes a more effective mechanism in the case of km-sized asteroids, capable of driving the rotation axis to the asymptotic state perpendicular to the orbital plane in just a few tens of millions of years.

# 5  Conclusions

The thermal inertia of an asteroid can be derived by comparing measurements of its thermal-infrared flux, at wavelengths typically between 5 and 20 μm, to synthetic fluxes generated by means of a thermophysical model (TPM). To date TPMs have been used to derive the thermal inertia of seven large MBAs and five NEAs. Although an extensive set of thermal-infrared observations of NEAs exists, application of TPMs is limited to the few asteroids for which the gross shape, the rotation period, and the spin axis orientation are known.

In order to overcome this limitation, we have developed a statistical method enabling the determination of the thermal inertia of a sample of objects for which such information is not available. This method has been applied to a sample of NEAs with diameters generally between 0.8 km and 3.4 km. The resulting value, $\Gamma$ = 200 ± 40 J m$^{-2}$ s$^{-0.5}$ K$^{-1}$, corresponds to a surface thermal conductivity of about 0.03 W m$^{-1}$ K$^{-1}$.

This value of thermal inertia and those derived by means of TPMs reveal a significant trend of increasing thermal inertia with decreasing asteroid diameter, *D*. Assuming that $\Gamma$ is proportional to $D^{-\xi}$ we derive a best-fit value for the exponent of $\xi \sim 0.4$.

The dependence Γ(*D*) has important implications for the magnitude of the Yarkovsky effect. On the basis of our results, the size dependence of the orbital semimajor axis drift rate $da/dt$ of MBAs for ~0.35 < *D* < ~10 due to the diurnal component of the Yarkovsky effect is proportional to $D^{-0.6}$, rather than the generally assumed $D^{-1}$ dependence for size-independent thermal inertia.

The modified dependence, $da/dt \propto D^{-0.6}$, implies that the size distribution of the objects injected by Yarkovsky-driven orbital mobility into the NEA source regions is less skewed to smaller sizes than generally assumed. This may help to explain the smaller-than-one difference in the value of the exponents of the cumulative size distribution of NEAs and MBAs.

We stress that the dataset on which our results are based is small and more multi-wavelength, multi-epoch thermal-infrared observations of asteroids with known spin states are required to refine our conclusions on the size dependence of thermal inertia and its consequences.

## 6 Acknowledgments

We wish to thank Bill Bottke and David Vokrouhlický for earlier suggestions and comments that inspired us to develop the original concept of this work further, and the referees of the present paper, Stephen Wolters and Bill Bottke, for suggestions that led to significant improvements in the presentation. M. D. wishes to acknowledge fruitful discussions with A. Morbidelli and A. Cellino.

# References

Allen, D. A. 1970. Infrared Diameter of Vesta. Nature 227, 158.

Bottke, W. F., Vokrouhlický, D., Rubincam, D. P., Nesvorný, D. 2006. The Yarkovsky and Yorp Effects: Implications for Asteroid Dynamics. Annual Review of Earth and Planetary Sciences 34, 157-191.

Bottke, W. F., Durda, D. D., Nesvorný, D., Jedicke, R., Morbidelli, A., Vokrouhlický, D., Levison, H. F. 2005. Linking the collisional history of the main asteroid belt to its dynamical excitation and depletion. Icarus 179, 63-94.

Bottke, W. F., Vokrouhlický, D., Rubincam, D. P., Broz, M. 2002. The Effect of Yarkovsky Thermal Forces on the Dynamical Evolution of Asteroids and Meteoroids. Asteroids III 395-408.

Bottke, W. F., Vokrouhlický, D., Broz, M., Nesvorný, D., Morbidelli, A. 2001. Dynamical Spreading of Asteroid Families by the Yarkovsky Effect. Science 294, 1693-1696.

Bottke, W. F., Jedicke, R., Morbidelli, A., Petit, J.-M., Gladman, B. 2000. Understanding the Distribution of Near-Earth Asteroids. Science 288, 2190-2194.

Britt, D. T., Yeomans, D., Housen, K., Consolmagno, G. 2002. Asteroid Density, Porosity, and Structure. Asteroids III 485-500.

Capek, D., Vokrouhlický, D. 2004. The YORP effect with finite thermal conductivity. Icarus 172, 526-536.

Chesley, S. R., Ostro, S. J., Vokrouhlický, D., Capek, D., Giorgini, J. D., Nolan, M. C., Margot, J.-L., Hine, A. A., Benner, L. A. M., Chamberlin, A. B. 2003. Direct Detection of the Yarkovsky Effect by Radar Ranging to Asteroid 6489 Golevka. Science 302, 1739-1742.

Christensen, P.R., and 21 colleagues. Morphology and composition of the surface of Mars: Mars Odyssey THEMIS results. Science 300, 2056-2061.

Davis, D. R., Gladman, B., Jedicke, R., Williams, G. 2006. The Sub-Kilometer Asteroid Diameter Survey. AAS/Division for Planetary Sciences Meeting Abstracts 38, #53.01.

Delbo', M., 2006. Delbo' Thermal Infrared Asteroid Diameters and Albedos V1.0. EAR-A-KECK1LWS/ETAL-5-DELBO-V1.0. NASA Planetary Data System, 2006." http://www.psi.edu/pds/resource/delbo.html

Delbo', M. 2004. The nature of near-earth asteroids from the study of their thermal infrared emission. Doctoral thesis, Freie Universität Berlin. http://www.diss.fu-berlin.de/2004/289/indexe.html.

Delbo', M., Harris, A. W., Binzel, R. P., Pravec, P., Davies, J. K. 2003. Keck observations of near-Earth asteroids in the thermal infrared. Icarus 166, 116-130.

Delbo', M., Harris, A. W. 2002. Physical properties of near-Earth asteroids from thermal infrared observations and thermal modeling. Meteoritics and Planetary Science 37, 1929-1936.

Domingue, D. L., Robinson, M., Carcich, B., Joseph, J., Thomas, P., Clark, B. E. 2002. Disk-Integrated Photometry of 433 Eros. Icarus 155, 205-219.

Emery, J. P., Sprague, A. L., Witteborn, F. C., Colwell, J. E., Kozlowski, R. W. H., Wooden, D. H. 1998. Mercury: Thermal Modeling and Mid-infrared (5-12 μm) Observations. Icarus 136, 104-123.

Farinella, P., Vokrouhlický, D., Hartmann, W. K. 1998. Meteorite Delivery via Yarkovsky Orbital Drift. Icarus 140, 369-378.

Hahn, G. 2006. Table on Physical Properties of NEOs. http://earn.dlr.de/nea/table1_new.html.

Hapke, B. 1984. Bidirectional reflectance spectroscopy 3. Correction for macroscopic roughness. Icarus 59, 41–59.

Harris, A. W., Mueller, M., Delbo', M., Bus, S. J. 2006. Physical Characterization of the Potentially-Hazardous High-Albedo Asteroid (33342) 1998 WT24 from Thermal-Infrared Observations. Icarus, in press.

Harris, A. W., 2006. The surface properties of small asteroids from thermal-infrared observations. In: Lazzaro, D., Ferraz-Mello, S., Fernández, J. A. (Eds.), Proc. of IAU Symposium 229. Cambridge University Press, Cambridge, UK, pp. 449 – 463.

Harris, A. W., Mueller, M., Delbo', M., Bus, S. J. 2005. The surface properties of small asteroids: Peculiar Betulia: a case study. Icarus, 179, 95-108.

Harris, A.W., and Lagerros, J.S.V. 2002. Asteroids in the thermal IR. In Bottke, W. F., Cellino, A., Paolicchi, P., Binzel, R. P. (Eds.), Asteroids III. Univ. of Arizona Press, Tucson, AZ, pp. 205 – 218.

Harris, A. W., Davies, J. K. 1999. Physical Characteristics of Near-Earth Asteroids from Thermal Infrared Spectrophotometry. Icarus 142, 464-475.

Harris, A. W., Davies, J. K., Green, S. F. 1998. Thermal Infrared Spectrophotometry of the Near-Earth Asteroids 2100 Ra-Shalom and 1991 EE. Icarus 135, 441-450.

Harris, A. W. 1998. A Thermal Model for Near-Earth Asteroids. Icarus 131, 291-301.

Ivezic, Z., and 31 coauthors. 2001. Solar System objects observed in the Sloan digital sky survey commissioning data. Astron. J. 122, 2749-2784.

Jakosky, B. M. 1986. On the thermal properties of Martian fines. Icarus 66, 117-124.

Keihm, S. J. 1984. Interpretation of the lunar microwave brightness temperature spectrum - Feasibility of orbital heat flow mapping. Icarus 60, 568-589.

Kührt, E., Giese, B. 1989. A thermal model of the Martian satellites. Icarus 81, 102-112.

Lagerros J. S. V. 1996. Thermal physics of asteroids I: Effects of shape, heat conduction and beaming. Astronomy and Astrophysics 310, 1011–1020.

La Spina, A., Paolicchi, P., Kryszczyńska, A., Pravec, P. 2004. Retrograde spins of near-Earth asteroids from the Yarkovsky effect. Nature 428, 400-401.

Lebofsky, L. A. and Spencer, J. R. 1989. Radiometry and thermal modeling of asteroids. In Asteroids II (R. P. Binzel, T. Gehrels, and M. S. Matthews, Eds.), University of Arizona Press, Tucson, 128-147.

Lim, L. F., McConnochie, T. H., Bell, J. F., Hayward, T. L. 2005. Thermal infrared (8-13 μm) spectra of 29 asteroids: the Cornell Mid-Infrared Asteroid Spectroscopy (MIDAS) Survey. Icarus 173, 385-408.

Morbidelli, A., Vokrouhlický, D. 2003. The Yarkovsky-driven origin of near-Earth asteroids. Icarus 163, 120-134.

Morbidelli, A., Bottke, W. F., Froeschlé, Ch., Michel, P., 2002. Origin and evolution of NEOs, in: Bottke, W.F., Cellino, A., Paolicchi, P., Binzel, R.P. (Eds.), Asteroids III, Univ. of Arizona Press, Tucson, pp. 409-422.

Mellon, M. T., Jakosky, B. M., Kieffer, H. H., Christensen, P. R. 2000. High-Resolution Thermal Inertia Mapping from the Mars Global Surveyor Thermal Emission Spectrometer. Icarus 148, 437-455.

Mueller, M., Delbo', M., Kaasalainen, M., Di Martino, M., Bus, S.J., Harris, A.W., 2006a. Indications for Regolith on Itokawa from Thermal-Infrared Observations. ASP Conference Series, in press

Mueller, M., Harris, A. W., Bus, S. J., Hora, J. L., Kassis, M., Adams, J. D. 2006b. The size and albedo of Rosetta fly-by target 21 Lutetia from new IRTF measurements and thermal modeling. Astronomy and Astrophysics 447, 1153-1158.

Müller, T. G., Blommaert, J. A. D. L. 2004. 65 Cybele in the thermal infrared: Multiple observations and thermophysical analysis. Astronomy and Astrophysics 418, 347-356.

Müller, T. G., Sekiguchi, T., Kaasalainen, M., Abe, M., Hasegawa, S. 2005. Thermal infrared observations of the Hayabusa spacecraft target asteroid 25143 Itokawa. Astronomy and Astrophysics 443, 347-355.

Müller, T. G., Sterzik, M. F., Schuetz, O., Pravec, P., Siebenmorgen, R. 2004. Thermal infrared observations of near-Earth asteroid 2002 NY40. Astronomy and Astrophysics 424, 1075-1080.

Müller, T. G., Lagerros, J. S. V. 1998. Asteroids as far-infrared photometric standards for ISOPHOT. Astronomy and Astrophysics 338, 340-352.

Mustard, J. F., Hays, J. E. 1997. Effects of Hyperfine Particles on Reflectance Spectra from 0.3 to 25 µm. Icarus 125, 145-163.

Nesvorný, D., Bottke, W. F. 2004. Detection of the Yarkovsky effect for main-belt asteroids. Icarus 170, 324-342.

Ostro, S.J. and 15 colleagues. 2004. Radar observations of asteroid 25143 Itokawa (1998 SF36). Meteoritics and Planetary Science 39, 407-424.

Pravec, P., Harris, A. W., Michalowski, T. 2002. Asteroid Rotations. Asteroids III 113-122.

Press, W. H., Teukolsky, S. A., Vetterling, W. T., Flannery, B. P. 1992. Numerical recipes in C. The art of scientific computing. $2^{nd}$ edition. Cambridge University Press, Cambridge / UK.

Putzig, N. E., Mellon, M. T., Kretke, K. A., Arvidson, R. E. 2005. Global thermal inertia and surface properties of Mars from the MGS mapping mission. Icarus 173, 325-341.

Rabinowitz, D.L., Helin, E., Lawrence, K., Pravdo, S., 2000. A reduced estimate of the number of kilometre-sized near-Earth asteroids. Nature 403, 165–166.

Rubincam, D. P. 2000. Radiative Spin-up and Spin-down of Small Asteroids. Icarus 148, 2-11.

Salisbury, J. W., D'Aria, D. M., Jarosewich, E. 1991. Midinfrared (2.5-13.5 microns) reflectance spectra of powdered stony meteorites. Icarus 92, 280-297.

Spencer, J. R. 1990. A rough-surface thermophysical model for airless planets. Icarus 83, 27-38.

Spencer, J. R., Lebofsky, L. A., Sykes, M. V. 1989. Systematic biases in radiometric diameter determinations. Icarus 78, 337-354.

Stuart, J. S., Binzel, R. P. 2004. Bias-corrected population, size distribution, and impact hazard for the near-Earth objects. Icarus 170, 295-311.

Veeder, G. J., Hanner, M. S., Matson, D. L., Tedesco, E. F., Lebofsky, L. A., Tokunaga, A. T. 1989. Radiometry of near-earth asteroids. Astronomical Journal 97, 1211-1219.

Vokrouhlický, D., Capek, D. 2002. YORP-Induced Long-Term Evolution of the Spin State of Small Asteroids and Meteoroids: Rubincam's Approximation. Icarus 159, 449-467.

Vokrouhlický, D. 1999. A complete linear model for the Yarkovsky thermal force on spherical asteroid fragments. Astronomy and Astrophysics 344, 362-366.

Wolters, S. D., Green, S. F., McBride, N., Davies, J. K. 2005. Optical and thermal infrared observations of six near-Earth asteroids in 2002. Icarus 175, 92-110.

Yoshida, F., Nakamura, T., Watanabe, J.-I., Kinoshita, D., Yamamoto, N., Fuse, T. 2003. Size and Spatial Distributionsof Sub-km Main-Belt Asteroids. Publications of the Astronomical Society of Japan 55, 701-715.

# Table 1

| Asteroid | | D (km) | $p_V$ | A | $\eta$ | r (AU) | α (°) | P (hrs) | Obs. Date | spin axis | re |
|---|---|---|---|---|---|---|---|---|---|---|---|
| 433 | Eros | 23.60 | 0.200 | 0.079 | 1.05 | 1.135 | 10 | 5.27 | 17-01-1975 | Y | a |
| 433 | Eros | 23.60 | 0.210 | 0.082 | 1.07 | 1.619 | 31 | 5.27 | 29-06-1998 | Y | a |
| 1580 | Betulia | 3.82 | 0.110 | 0.043 | 1.09 | 1.199 | 53 | 6.138 | 22-06-2002 | Y | b |
| 1862 | Apollo | 1.40 | 0.260 | 0.102 | 1.15 | 1.063 | 35 | 3.065 | 26-11-1980 | - | c |
| 1866 | Sisyphus | 8.90 | 0.140 | 0.055 | 1.14 | 1.609 | 35 | 2.4 | 29-06-1998 | - | d |
| 1980 | Tezcatlipoca | 6.60 | 0.150 | 0.059 | 1.64 | 1.129 | 63 | 7.252 | 31-08-1997 | Y | a |
| 2100 | Ra-Shalom | 2.79 | 0.080 | 0.031 | 2.32 | 1.174 | 39 | 19.8 | 21-08-2000 | Y | e |
| 2100 | Ra-Shalom | 2.50 | 0.130 | 0.051 | 1.80 | 1.195 | 41 | 19.8 | 30-08-1997 | Y | f |
| 3200 | Phaethon | 5.10 | 0.110 | 0.043 | 1.60 | 1.132 | 48 | 3.604 | 20-12-1984 | - | c |
| 3554 | Amun | 2.10 | 0.170 | 0.067 | 1.20 | 1.243 | 16 | 2.53 | 15-03-1986 | - | c |
| 3671 | Dionysus | 1.50 | 0.160 | 0.063 | 3.10 | 1.126 | 58 | 2.705 | 02-06-1997 | - | a |
| 5381 | Sekhmet | 1.50 | 0.220 | 0.086 | 1.90 | 1.213 | 44 | 3 | 22-06-2003 | - | d |
| 5381 | Sekhmet | 1.40 | 0.240 | 0.094 | 1.50 | 1.119 | 35 | 3 | 14-05-2003 | - | d |
| 5587 | 1990 SB | 4.00 | 0.250 | 0.098 | 1.10 | 1.399 | 19 | 5.052 | 09-04-2001 | Y | d |
| 5587 | 1990 SB | 3.57 | 0.320 | 0.126 | 0.84 | 1.210 | 42 | 5.052 | 10-05-2001 | Y | e |
| 6455 | 1992 HE | 3.43 | 0.280 | 0.110 | 0.80 | 1.641 | 22 | - | 22-03-2002 | - | g |
| 6455 | 1992 HE | 3.55 | 0.240 | 0.094 | 0.70 | 1.364 | 29 | - | 29-09-2002 | - | g |
| 9856 | 1991 EE | 1.00 | 0.300 | 0.118 | 1.15 | 1.093 | 36 | 3.045 | 11-09-1991 | - | f |
| 14402 | 1991 DB | 0.60 | 0.140 | 0.055 | 1.04 | 1.025 | 36 | 2.266 | 16-04-2000 | - | e |
| 19356 | 1997 $GH_3$ | 0.91 | 0.340 | 0.133 | 0.98 | 1.406 | 5 | 6.714 | 11-05-2001 | - | e |
| 25330 | 1999 $KV_4$ | 2.55 | 0.080 | 0.031 | 1.06 | 1.392 | 3 | 4.919 | 14-05-2003 | - | d |
| 25330 | 1999 $KV_4$ | 2.70 | 0.080 | 0.031 | 1.30 | 1.495 | 16 | 4.919 | 02-06-2003 | - | d |
| 25330 | 1999 $KV_4$ | 3.21 | 0.050 | 0.020 | 1.50 | 1.197 | 54 | 4.919 | 10-05-2001 | - | e |
| 33342 | 1998 $WT_{24}$ | 0.34 | 0.590 | 0.232 | 0.90 | 0.990 | 67 | 3.697 | 18-12-2001 | - | d |
| 33342 | 1998 $WT_{24}$ | 0.44 | 0.350 | 0.137 | 1.50 | 0.987 | 79 | 3.697 | 19-12-2001 | - | d |
| 33342 | 1998 $WT_{24}$ | 0.50 | 0.270 | 0.106 | 1.85 | 0.981 | 93 | 3.697 | 21-12-2001 | - | d |
| 35396 | 1997 $XF_{11}$ | 0.89 | 0.320 | 0.126 | 1.30 | 1.215 | 30 | 3.257 | 28-11-2002 | - | d |
| 35396 | 1997 $XF_{11}$ | 0.91 | 0.310 | 0.122 | 1.20 | 1.018 | 53 | 3.257 | 03-11-2002 | - | d |
| 35396 | 1997 $XF_{11}$ | 1.18 | 0.180 | 0.071 | 1.80 | 1.034 | 63 | 3.257 | 05-11-2002 | - | d |
| 53789 | 2000 $ED_{104}$ | 1.20 | 0.180 | 0.071 | 1.68 | 1.089 | 60 | - | 29-09-2002 | - | g |
| 85953 | 1999 $FK_{21}$ | 0.59 | 0.320 | 0.126 | 0.91 | 1.140 | 35 | - | 21-02-2002 | - | e |
| 86039 | 1999 $NC_{43}$ | 2.22 | 0.140 | 0.055 | 2.86 | 1.116 | 59 | 34.49 | 17-03-2000 | - | e |
| 99935 | 2002 $AV_4$ | 1.50 | 0.370 | 0.145 | 1.60 | 1.086 | 70 | - | 01-06-2003 | - | d |
| | 1999 $HF_1$ | 4.74 | 0.110 | 0.043 | 1.68 | 0.957 | 91 | - | 22-03-2002 | - | g |
| | 2000 $BG_{19}$ | 1.77 | 0.040 | 0.016 | 0.74 | 1.388 | 17 | - | 17-03-2000 | - | e |
| | 2001 LF | 2.00 | 0.050 | 0.020 | 1.40 | 1.172 | 45 | - | 02-06-2003 | - | d |
| | 2002 $BM_{26}$ | 0.84 | 0.020 | 0.008 | 3.10 | 1.023 | 60 | - | 21-02-2002 | - | e |
| | 2002 $HK_{12}$ | 0.80 | 0.170 | 0.067 | 2.84 | 1.138 | 33 | - | 28-09-2002 | - | g |
| | 2002 $NX_{18}$ | 2.40 | 0.030 | 0.012 | 1.19 | 1.158 | 54 | - | 29-09-2002 | - | g |
| | 2002 $QE_{15}$ | 1.94 | 0.150 | 0.059 | 1.53 | 1.131 | 62 | - | 28-09-2002 | - | g |
| | 2003 $YT_1$ | 1.50 | 0.270 | 0.106 | 1.92 | 1.035 | 74 | - | 08-05-2004 | - | d |

Table 1 Near Earth-asteroids with η-values derived from spectral fitting to multi-wavelength mid-infrared observations. The object effective diameter, *D*, the geometric visible albedo $p_V$, and the η-values have been derived by using the NEATM. α is the phase angle at the epoch of the observations, which is given in the "Obs. Date" column. *P* is the rotation period in hours. In the column "Spin axis" a "Y" indicates that the spin axis orientation of the asteroid is known. In the column "Re" we give the original publication reference: a) Harris and Davies (1999); b) Harris *et al.* (2005); c) Harris (1998); d) Delbo' (2004); e) Delbo' *et al.* (2003); f) Harris *et al.* (1998); g) Wolters *et al.* (2005).

## Figure captions

Fig. 1 Sensitivity of η to model parameter variations. Δη/η (%) caused by a change of 1% in the bolometric Bond albedo *A* (upper left), macroscopic surface roughness $\bar{\theta}_0$ (upper right), heliocentric distance *r* (lower left), and the thermal parameter Θ (lower right). See text, section 2.1 for details.

Fig. 2 Distribution of the input parameters used in our statistical inversion method. Upper left: distributions of NEA rotation rates from "Physical parameters of NEOs (Hahn, 2006, http://earn.dlr.de"). Upper right: distribution of NEA phase angles. Lower left: distribution of NEA sub-solar latitudes $\theta_{SS}$. Lower right, solid line: distribution of the geometric albedos ($p_V$) and, dashed line: the bolometric Bond albedos (*A*) for the asteroids of Table 1 having η-values determined from observations.

Fig. 3 Dependence of η-value on phase angle, α. Black diamonds: η-values derived from the NEATM for a set of NEAs with adequate multi-filter photometric data to enable η to be derived via spectral fitting (the data set includes multiple values of η for some objects observed at more than one phase angle; for the original data sources see Table 1). The error bars represent a 20% uncertainty, which is based on the reproducibility of η for those objects for which more than one measurement is available from independent data sets. Colored points: distributions of (α, η) calculated by means of our model for different values of thermal inertia: i.e. 15 (green), 200 (red), and 1000 (blue) J m$^{-2}$ s$^{-0.5}$ K$^{-1}$. The distribution of the measured η-values is best described by the red points.

Fig. 4 Plot of the function *D* (Γ), i.e. the distance D of the two-dimensional Kolmogorov-Smirnov best-fit procedure against the thermal inertia Γ. The three curves were generated assuming three different values of the surface roughness: solid line $\bar{\theta}=58°$; dotted line $\bar{\theta}=36°$; dashed line $\bar{\theta}=0°$ i.e. a smooth surface. The dashed-dotted line shows the function *D* (Γ) obtained by using $\bar{\theta}=58°$ and a random distribution of asteroid spin-axis orientations uniformly distributed over the celestial sphere, instead of the nominal one, as input for our model.

Fig. 5 Histogram of the distribution of the log of the diameters, *D*, of the NEAs for which we have η-values determined from observations. The best-fit Gaussian function, $0.37\exp(-z^2/2)$, where $z=(\log D - 0.23)/0.31$, with *D* in km, is also shown.

Fig. 6 Thermal inertia as a function of asteroid diameter. Small open circles represent values from the literature derived by means of thermophysical models. The large open <u>diamond</u> is the result from this work (see text for details). The straight (continuous) line which gives the best fit to the trend of increasing thermal inertia, Γ, with decreasing asteroid diameter, *D*, is given by the expression $\Gamma=300\times D^{-0.48}$. The axis on the right-hand side gives the asteroid surface thermal conductivity *k* on the basis of $k=\Gamma^2/(\rho c)$, assuming constant surface density, ρ, equal to 2500 kg m$^{-3}$ and specific heat capacity, *c*, equal to 600 J kg$^{-1}$ K$^{-1}$. These values are reasonable assumptions for asteroid surfaces (Britt *et al.*, 2002; Farinella *et al.*, 1998). The thermal conductivities of (6489) Golevka (Chesley *et al.*, 2003) and for Karin cluster asteroids (Nesvorný and Bottke, 2004) are indicated with arrows. The two values of Γ derived for 2002 NY$_{40}$ are indicated as the lower and the upper limits of the error bar on the extreme left of the plot. Dotted line: linear regression of Eq. (6) for MBAs only; dashed line: linear regression of Eq. (6) for NEAs only.

# Figures

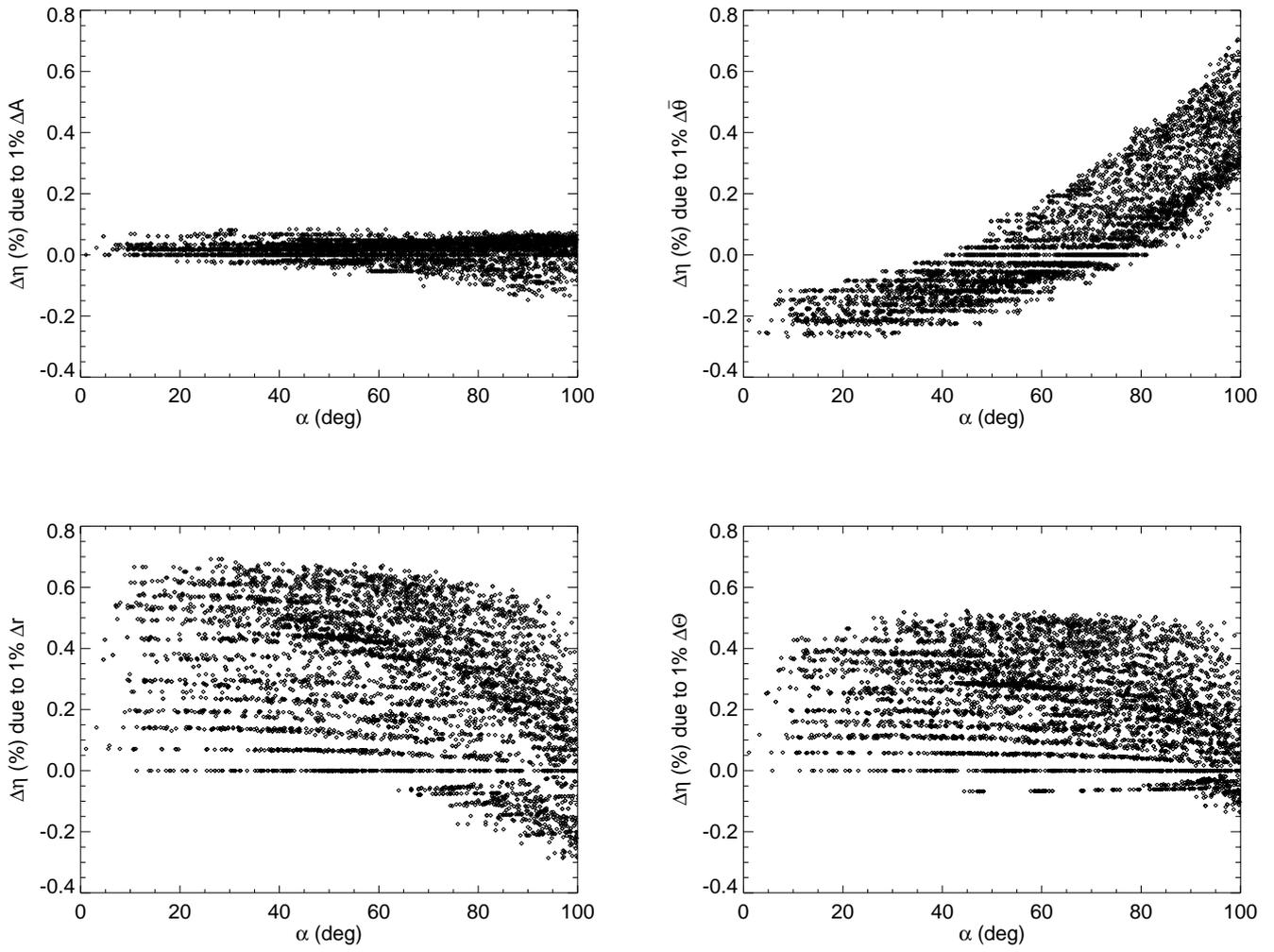

Fig. 1

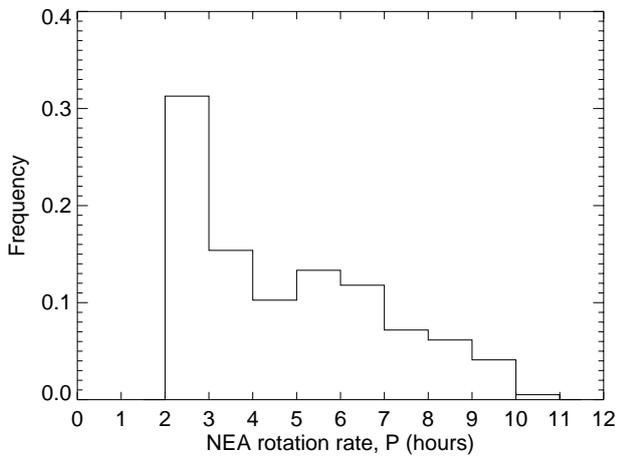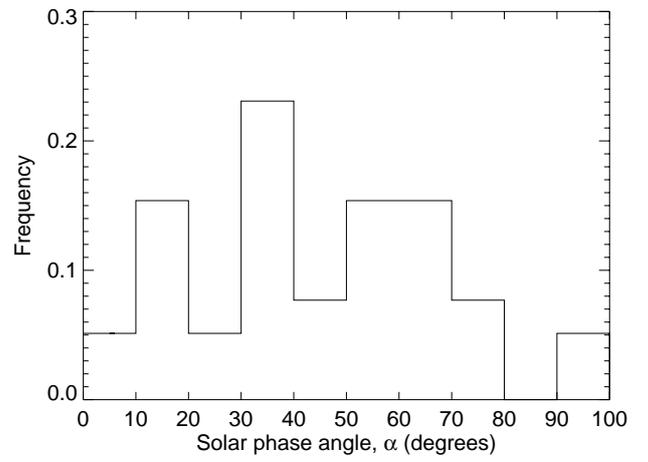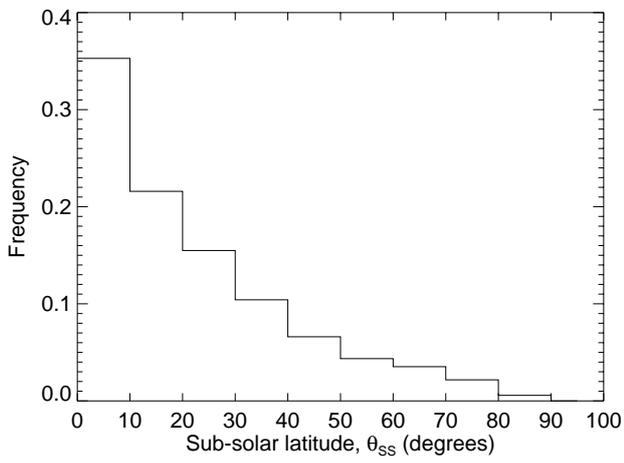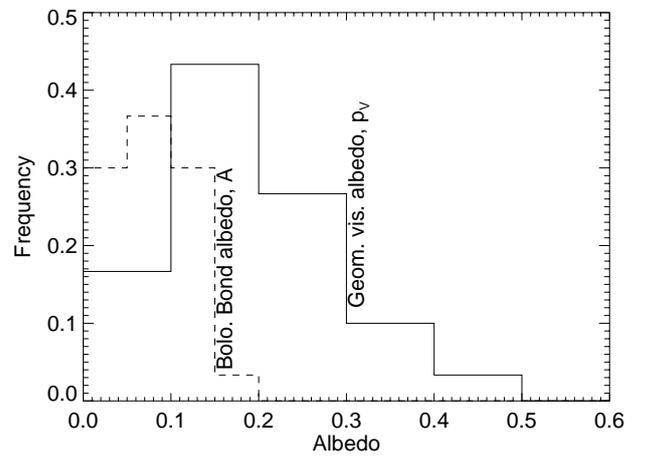

Fig. 2

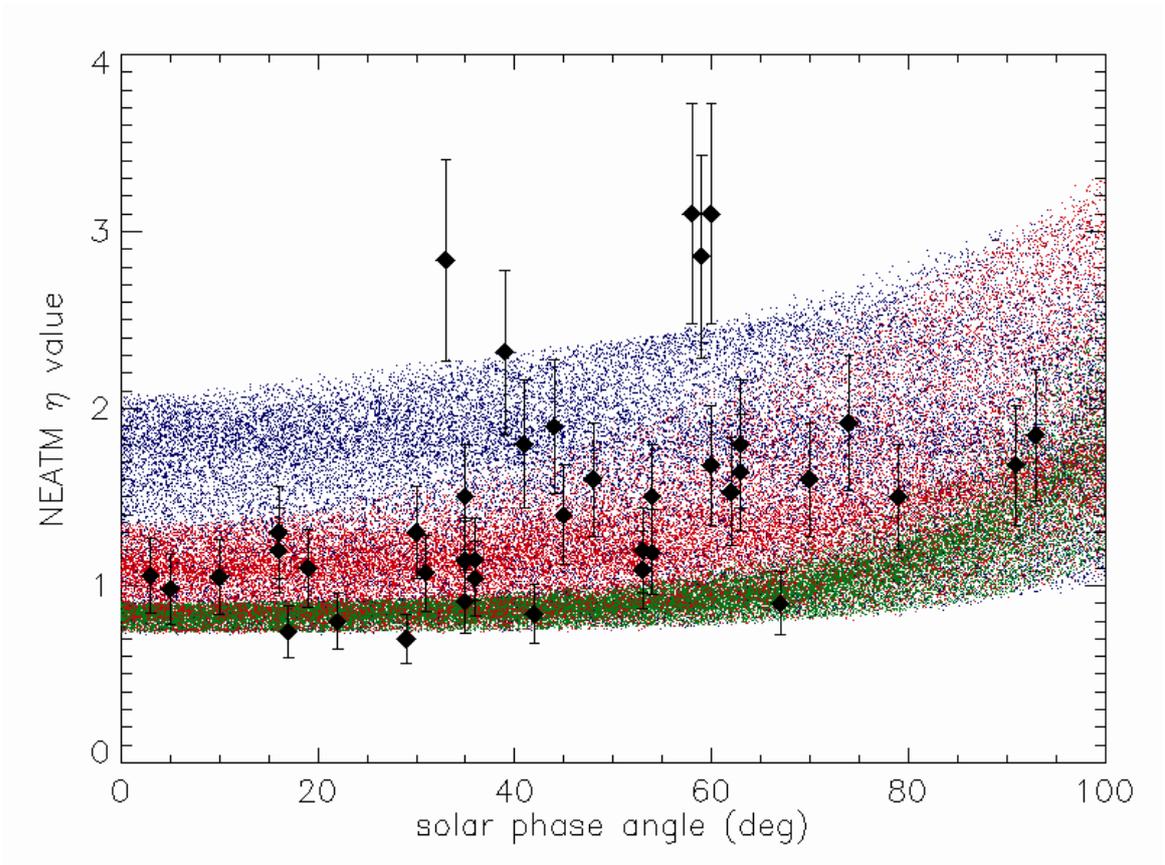

Fig. 3

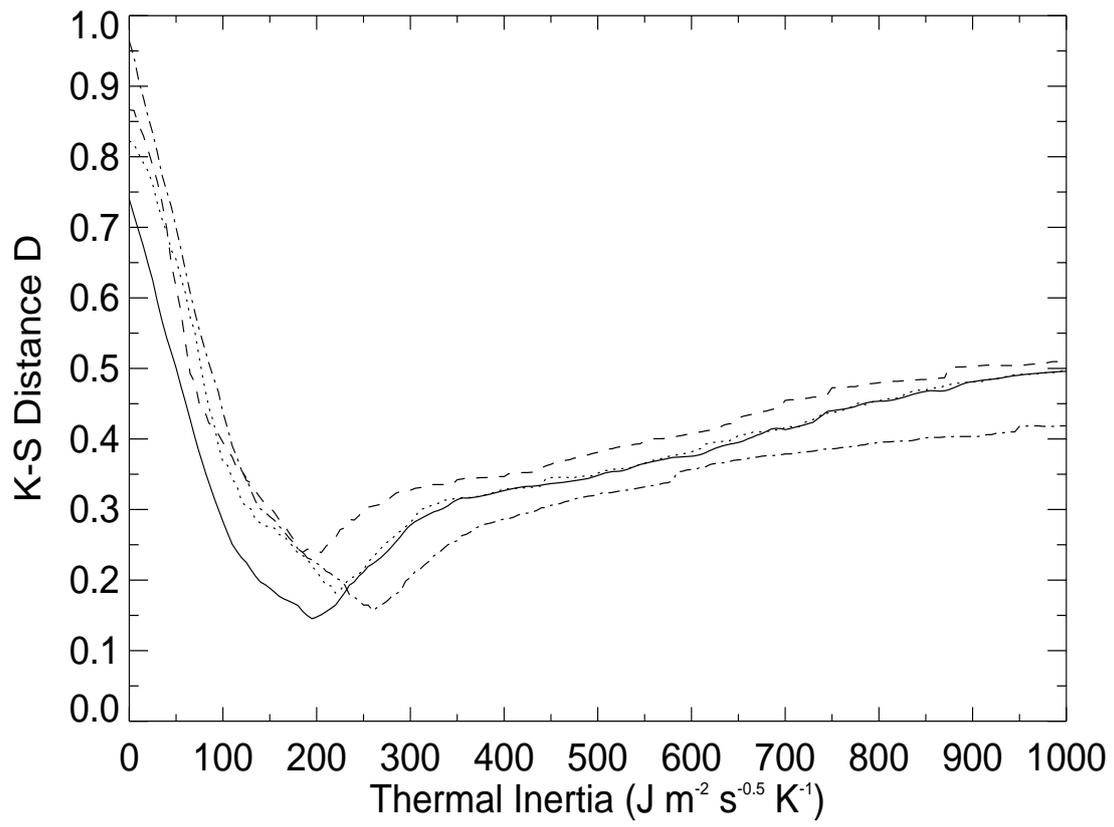

Fig. 4

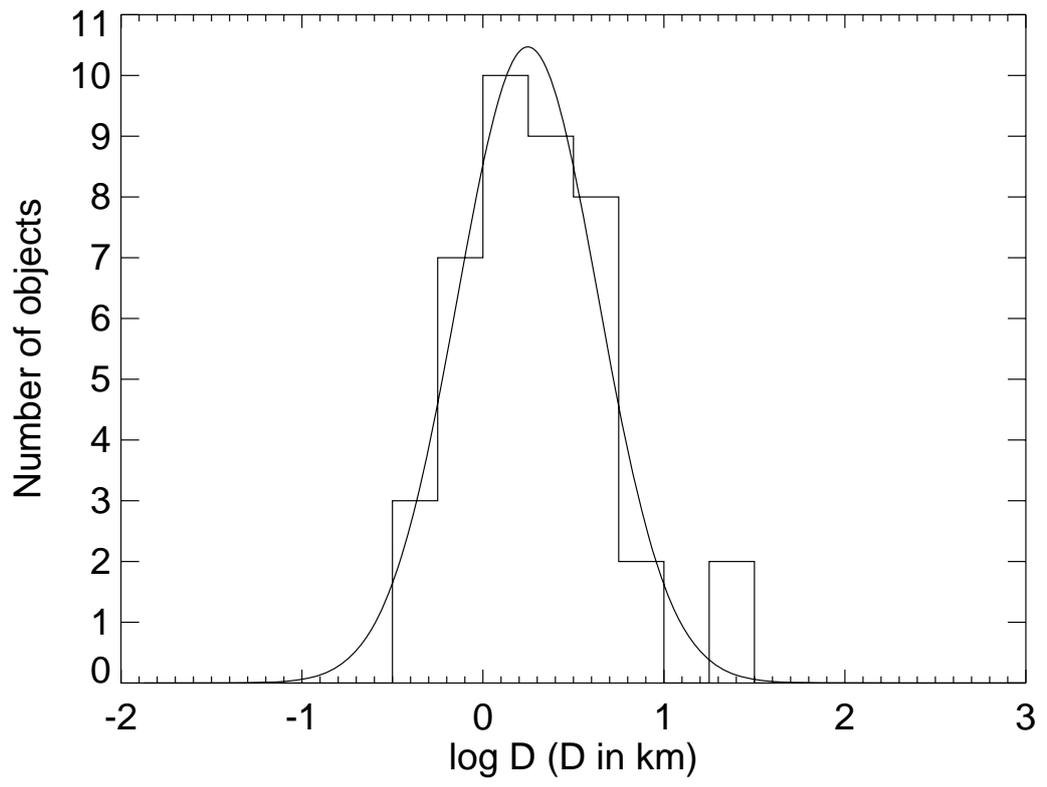

Fig. 5

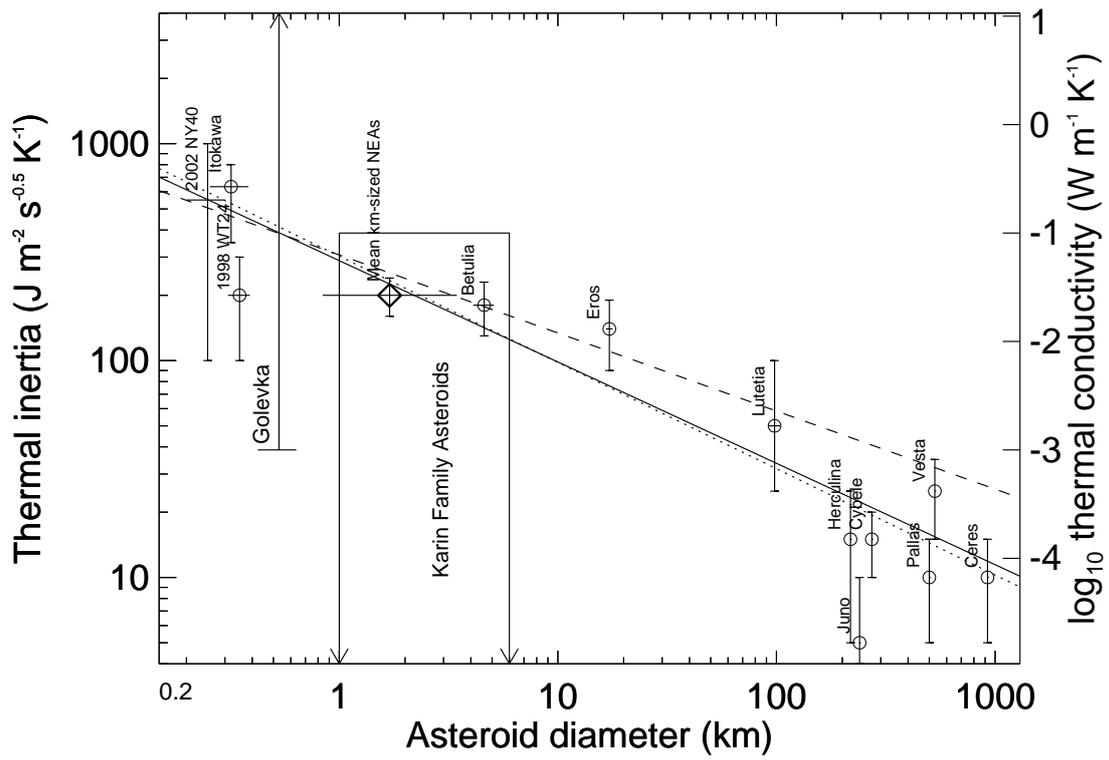

Fig. 6